# Mid-infrared Active Graphene Nanoribbon Plasmonic Waveguide Devices

**Kelvin J. A. Ooi,**[1,2,3] **Hong Son Chu,**[1,4] **Lay Kee Ang,**[2,3] **Ping Bai,**[1,5]

[1]*Electronics & Photonics Department, A*STAR Institute of High Performance Computing, 1 Fusionopolis Way, #16-16 Connexis, Singapore 138632*

[2]*School of Electrical and Electronic Engineering, Nanyang Technological University, Block S1, 50 Nanyang Avenue, Singapore 639798*

[3]*Engineering Product Development, Singapore University of Technology and Design, 20 Dover Drive, Singapore 138682*

[4]*Corresponding author: chuhs@ihpc.a-star.edu.sg*

[5]*Corresponding author: baiping@ihpc.a-star.edu.sg*

**Abstract**

Doped graphene emerges as a strong contender for active plasmonic material in the mid-infrared wavelengths due to the versatile external-control of its permittivity-function and also its highly-compressed graphene surface plasmon (GSP) wavelength. In this paper, we design active plasmonic waveguide devices based on electrical-modulation of doped graphene nanoribbons (GNRs) on a voltage-gated inhomogeneous dielectric layer. We first develop figure-of-merit (FoM) formulae to characterize the performance of passive and active graphene nanoribbon waveguides. Based on the FoMs, we choose optimal GNRs to build a plasmonic shutter, which consists of a GNR placed on top of an inhomogeneous $SiO_2$ substrate supported by a Si nanopillar. Simulation studies show that for a simple 50nm-long plasmonic shutter, the modulation contrast can exceed 30dB. The plasmonic shutter is further extended to build a 4–port active power splitter and an 8-port active network, both based on GNR cross-junction waveguides.  For the active power splitter, the GSP power transmission at each waveguide arm can be independently controlled by an applied gate-voltage with high modulation contrast and nearly-equal power-splitting proportions. From the construct of the 8-port active network, we see that it is possible to scale up the GNR cross-junction waveguides into large and complex active waveguide networks, showing great potential in an exciting new area of mid-infrared graphene plasmonic integrated nanocircuits.

## 1.    Introduction

There is great scientific and technological interest in the mid-infrared spectral range of 2–20 μm recently. This spectral range enables several applications in photonics such as spectroscopy, materials processing, chemical and biomolecular sensing, remote explosive detection, and covert communication systems [1]. Moreover, the mid-infrared spectral region is also attractive for the study of plasmonic devices, opening up possibilities of integration with electronic devices designing devices with active control over the surface plasmon resonance at metal/dielectric interfaces [2–5]. However, due to the metal's relatively weak refractive index change with electrical-bias, mechanical force, or temperature, the active control of metal-based plasmonics is challenging [6]. As such, most of the current active metal-based plasmonic devices require an active dielectric medium [7, 8], which are often inefficient due to their weak nonlinear material properties, leading to the necessity of dimensionally-long devices. Recently, graphene, a two-dimensional material with one atomic thickness, is a promising candidate for active plasmonic material [9]. In the mid-infared wavelengths, when graphene is doped with a certain electron or hole concentration, it is demonstrated both theoretically [10–13] and experimentally [14–16] that graphene surface plasmons (GSP) can exist. Graphene is an excellent platform for plasmonic devices owing to its large active-control of its permittivity-function that is not seen in noble metals [17]. Furthermore, graphene plasmonic devices are very compact due to the highly-compressed GSP wavelength [9, 12], and demonstrates active waveguide-based switching [18–20], as well as low sharp-bend losses [21], which could lead to an exciting new area of research in mid-infrared graphene plasmonic nanocircuits. In this paper, we first investigate the propagation characteristics of doped graphene sheets and nanoribbons to determine the best GSP parameters to be used for the construction of the active waveguides. We will then use these GSP parameters to design ultracompact and high-contrast active waveguide devices, including plasmonic shutters, active power splitters and active plasmonic waveguide networks.

**2. Propagation characteristics of doped graphene sheets and nanoribbon waveguides**

*2.1. Graphene sheets*

To describe the optical properties of the graphene monolayer, we first obtain the dynamic conductivity of the graphene sheet, σ(ω), which is derived from the Kubo formula [10,11]. It consists of an intraband contribution:

$$\sigma_{intra}(\omega) = \frac{ie^2 \mu}{\pi \hbar^2 (\omega + i\tau^{-1})} \qquad (1)$$

and an interband contribution:

$$\sigma_{inter}(\omega) = \frac{ie^2}{4\pi\hbar} \ln\left( \frac{2|\mu| - (\omega + i\tau^{-1})\hbar}{2|\mu| + (\omega + i\tau^{-1})\hbar} \right) \qquad (2)$$

under the condition $k_B T \ll |\mu|$, where i is the imaginary unit, e is the charge of the electron, μ is the chemical potential, ℏ is the reduced Planck's constant, ω is the radian frequency, τ=0.5ps is the relaxation time, $k_B$ is the Boltzmann constant, and T is the temperature. From here we can deduce the existence of two modes: the transverse-magnetic (TM) GSP mode, which is supported when μ>0 and **Im**($\sigma_{intra}$)>0; and the transverse-electric (TE) mode, which is supported when **Im**($\sigma_{inter}$)<0 and |μ|<ℏω/2 [10]. It is noted from equations (1) and (2) that by controlling μ, we would be able to change the optical conductivity of graphene. As a result, it enables switching between the supported and unsupported GSP modes in a graphene sheet. μ can either be controlled by chemical doping [22], or via an electric-field according to the equation [17,23]:

$$\mu = \hbar v_F \sqrt{\frac{\pi \varepsilon \varepsilon_0}{e} E} \qquad (3)$$

where $v_F=10^6$ m/s is the Fermi velocity, $\varepsilon$ is the dielectric spacer relative permittivity, $\varepsilon_0$ is the free space permittivity, and E is the electric-field.

The obtained graphene conductivity can be used to define the GSP wave vector:

$$\beta^2 = k_0^2 \left[1 - \left(\frac{2}{\eta_0 \sigma}\right)^2\right] \quad (4)$$

where $k_0$ is the free space wave vector and $\eta_0=377\Omega$ is the intrinsic impedance of free space. This parameter is then used to define the GSP wavelength, $\lambda_{GSP}=2\pi/\mathbf{Re}(\beta)$, and also the GSP propagation length, $L_{GSP}=1/\mathbf{Im}(\beta)$. Taken together, the Figure-of-Merit (FoM) for GSP propagation on a graphene sheet can be defined, which is given as [13]:

$$\text{FoM}_1 = \frac{L_{GSP}}{\sqrt{\lambda_{GSP} \cdot \lambda_0}} \quad (5a)$$

FoM$_1$ is plotted in Figure 1(a) over a spectrum of free space wavelength $\lambda_0$ and chemical potential $\mu$. It is noticed that the best-performing FoM$_1$ values generally occur at short $\lambda_0$ with high $\mu$. However, this does not give a good picture of the tradeoffs in constructing such highly-doped graphene, given the complexity of achieving high chemical doping [24], or the enormous electric-field needed to achieve such high $\mu$ as presented in equation (3). As such, we offer an improved FoM definition which is normalized by the chemical potential and given as:

$$\text{FoM}_2 = \text{FoM}_1 / \mu[\text{eV}] \quad (5b)$$

FoM$_2$ as a function of $\lambda_0$ and $\mu$ is plotted in Figure 1(b). Indeed, FoM$_2$ shows that the performance of GSP propagation at short $\lambda_0$ and high $\mu$ is on par with those at long $\lambda_0$ and low $\mu$. As such, one has a choice of designing low-doped or low-powered graphene plasmonic devices in the mid-infrared regime.

The two FoMs presented above are for passive waveguides. For active waveguides, we need to take into account the switching performance of the graphene. Similar to most optical modulators, there are two modulation schemes for graphene as well, namely, the phase modulation scheme, with FoM given as:

$$\text{FoM}_3 = \text{FoM}_2 \times \frac{\Delta\lambda_{GSP}}{\lambda_{GSP}} \quad (5c)$$

and the absorption modulation scheme, with FOM given as:

$$\text{FoM}_4 = \text{FoM}_2 \times \frac{\Delta L_{GSP}}{L_{GSP}} \quad (5d)$$

Both FoM$_3$ and FoM$_4$ are plotted in Figures 1(c) and 1(d) respectively. It is observed that the switching performance for both schemes is particularly good in the low $\mu$ regime. Considering the FoMs for both passive and active waveguides, the best waveguide performance occurs for operating wavelengths between 6–10μm and μ below 0.3eV. Here, we will be using $\lambda_0$=10μm and μ=0.2eV as the reference operating wavelength and chemical potential for all the graphene plasmonic devices discussed here.

*2.2. Graphene nanoribbons*

The GSP propagation on finite width graphene nanoribbons (GNRs) is very different from those on infinite graphene sheets. For example, a $\lambda_0$=10μm GSP propagating on a μ=0.2eV doped free-standing graphene sheet yields $\lambda_{GSP}$≈200nm and $L_{GSP}$≈2.6μm. However, results in Figure 2(a) obtained from mode simulation using C. S. T. Microwave Studio show that $\lambda_{GSP}$ and $L_{GSP}$ strongly depend on the ribbon width and significantly differ from their graphene sheet values. Also, due to the emergence of edge modes – as described in detail in refs. [25, 26] – the fundamental propagation mode is no longer a pure TM mode, but now a quasi-TEM mode. This is shown in Figures 2(c) and (d), whereby the propagation mode of a 30nm-wide GNR consists of both a z-direction and y-direction electric-field component. At very narrow GNR widths, $\lambda_{GSP}$ is further compressed down by more than 2 times due to strong coupling between the edge modes. Therefore, the FoMs are also width-dependent, as observed in Figure 2(b). Both passive and active waveguides have good FoMs in the GNR width range from 30–50nm. For the subsequent designs of active GNR waveguides, we will adopt a GNR width of 30nm throughout. $\lambda_{GSP}$ and $L_{GSP}$ for this freestanding GNR are ~120nm and ~1.9μm respectively.

## 3. Design of graphene plasmonic devices

We designed the graphene plasmonic devices by 3D full-wave simulation as follows using C. S. T. Microwave Studio. However, since graphene is a 2D material, we could not directly apply the graphene parameters (such as surface conductivity and GSP wave vector) into the software. Instead, we employed a special technique to model graphene in 3D simulation software environment, which was first introduced by Vakil and Engheta [12]. By defining a certain thickness Δ of graphene, (e.g. Δ = 1nm), the "pseudo" bulk-conductivity can be defined as σ(ω)/Δ, and thus the "pseudo" permittivity can also be defined as ε(ω) = 1+iσ(ω)/($\varepsilon_0$ωΔ). The permittivity values can thus be used as material parameters in C. S. T. Microwave Studio. The obtained permittivity values were used strictly in conjunction with the defined thickness value. Using this technique, we found that our simulation results for graphene sheets matched well with the results from analytical formula.

*3.1. Graphene plasmonic shutter*

We first design a graphene plasmonic shutter, which consists of a graphene nanoribbon (GNR) placed on top of an inhomogeneous silicon dioxide ($SiO_2$) dielectric layer [12]. The schematic in Figure 3 shows that the GNR is placed on a silicon-on-insulator (SOI) substrate, where a silicon (Si) nanopillar of 30nm×50nm×100nm is located at the middle section and surrounded by $SiO_2$ cladding medium. The GNR is separated by a 10nm gate-oxide gap from the Si nanopillar. The GNR can be chemically doped to the chemical potential level of μ=0.2eV [22], so that under an unbiased condition, GSP can propagate through the waveguide. Since the GNR is not free-standing and sits on $SiO_2$/Si (ε=2.1/11.7 [27]) stack, $\lambda_{GSP}$ is further scaled down to ~80nm.

During the modulation process, we apply a gate-voltage to reduce the chemical potential of graphene to the propagation cutoff. The intensity plot in Figure 3 obtained from simulation shows that the application of a bias voltage on a 50nm-long ($<\lambda_{GSP}$) GNR (e.g. biasing to µ=0.05eV and hence ∆µ=−0.15eV or E=0.143V/nm) can modulate the GSP transmission by >30dB contrast. The obtained result is consistent with the graphene plasmonic modulator reported in ref. [28].

We note that the presence of the Si nanopillar just 10nm below the GNR layer causes an abrupt change in the effective refractive index at the middle of the waveguide. In this portion of the GNR waveguide, the GSP is more confined compared to the rest of the waveguide, and hence the intensity is higher in this region as observed from the intensity plot for the unbiased-state. This also induces a slight reflection loss of 0.3dB. Increasing the gate-oxide gap thickness might reduce this effect, but at the expense of increasing the applied gate-voltage.

In the biased-state, because the middle section of the GNR is at the propagation cutoff state, there is a large reflection of 2dB induced, which can be seen from the intensity peak from Figure 3 at x=125nm, right at the interface between two GNRs with different chemical potential levels. This reflection, if not circumvented, would greatly affect the performance of the GNR power splitter, which will be discussed in the next section.

*3.2.     Active power splitter based on graphene nanoribbon cross-junction waveguides*

In this section, we further expand the idea of the plasmonic shutter to build a 4–port active power splitter based on a GNR cross-junction waveguide [29]. The 4–port active power splitter is illustrated in Figure 4(a). GSP entering the input port (Port 1) would be split evenly into the three output ports. The Si nanopillars, each having a dimension of 30nm×50nm×100nm, are electrically isolated at the junction so that the cross-junction waveguide arms can be individually controlled, as shown in Figure 4(b). The separation distance between two adjacent Si nanopillars consists of a total buffer and junction length of 2×$L_b$+30nm. In this proposed active power splitter, the GNR is doped to µ=0.2eV to enable the GSP propagation in the unbiased state. By applying an appropriate gate-voltage across an individual waveguide arm, the chemical potential is reduced to 0.05eV and thus the waveguide arm transmission is switched to cutoff.

The active waveguide arms are not directly connected to the junction. Instead, they are separated by a buffer zone, which has a buffer length $L_b$ from the junction. The buffer zones are required for optimal GSP power-splitting during the modulation of the waveguide arms. It reduces the reflection and minimizes the imbalance between the splitting proportions for the transmitting outputs. When one or more of the waveguide arms are switched off, there is still a slight protrusion of $L_b$ from the junction that remains transmitting and hence forms a stub-like structure, as is illustrated in the z-direction electric-field maps in Figures 5(a)–(e). We have enlarged the electric-field map of Figure 5(e) into Figure 5(f) to clearly show the formation of the stub-like structures when two waveguide arms are switched off. These plasmonic stub-like structures optimize the phase-matching to reduce reflections and maximize the power transmission over the other output arms when one or more waveguide arms are switched off [30, 31].

In Figures 5(a)–(e), we also show the transmitted power at the output ports as a function of buffer length $L_b$ with respect to 5 cases, namely (a) all-transmitting output ports, (b) transmitting port 2 only, (c) transmitting ports 2 and 3, (d) transmitting ports 3 and 4, and (e) transmitting port 3 only. It is noticed from the power distribution plots that without an optimized buffer zone, switching off individual or multiple waveguide arms would, in some situations, unbalance the power splitting proportion among the other waveguide arms, and also reduce the transmission intensity as a result of reflection. Analyzing the results plotted in Figure 5, the best buffer lengths $L_b$ occur from 15–35nm (shaded region), whereby most of the output power transmission intensities converge and are at maximum.

*3.3.    Cross-junction active waveguide networks*

The cross-junction waveguide can be further scaled up to a large array of waveguide networks [32]. We show a few examples of GSP routing in an 8–port network in Figures 6(a)–(c), which consists of a 2×2 square array of cross-junction waveguides. Similar to the modulation scheme as aforementioned, the GNR is doped to µ=0.2eV to allow GSP propagation in the unbiased state. Then, applying gate-voltages to individual arms, the chemical potential is lowered to 0.05eV and hence GSP transmission is switched to cutoff. This enables the routing of GSP from the input port to the specific output port with very good modulation contrast. The active GNR waveguide network opens a potential possibility to design active and ultracompact cascading power splitters, multiplexers and demultiplexers for multichannel communications.

## 4.    Conclusion

We have built active plasmonic waveguide devices, including plasmonic shutters and active cross-junction waveguide networks, with doped graphene nanoribbons. FoMs are developed to evaluate the optical performance of passive and active graphene sheets and nanoribbon waveguides. Based on the FoMs, we chose 10µm as a suitable operating wavelength, a 30nm GNR width and 0.2eV chemical doping level to build active GNR waveguide devices. We first design a GNR plasmonic shutter, where the modulation of a 50nm-long (<$\lambda_{GSP}$) GNR on top of a 30nm×50nm×100nm Si nanopillar exhibits a modulation contrast exceeding 30dB. The concept of the plasmonic shutter is used in the design of a 4–port active GNR cross-junction waveguide power splitter, where each waveguide arm can be independently switched on or off with a high modulation contrast. The power splitter shows nearly-equal power-splitting proportions even during modulation, which is achieved by the use of plasmonic stub-like structures for phase-matching. The cross-junction waveguide can be further scaled up into an array of GNR waveguide networks, which, for example, can be used as an 8–port active cascaded power splitter, with independent control for each of the 8 waveguide arms. Owing to the highly-compressed $\lambda_{GSP}$, the GNR waveguide networks can be made ultracompact, showing limitless potential in scalability, complexity and integrability for future nanocircuitry and nanodevices. The proposed GNR waveguide devices and structures open up an exciting field of mid-infrared graphene plasmonic integrated nanocircuits operating in deep sub-micron dimensions.

## 5.    Acknowledgments

This work was supported by the Agency for Science and Technology Research (A*STAR), Singapore, Metamaterials-Nanoplasmonics research programme under A*STAR-SERC grant No. 0921540098 and the National Research Foundation Singapore under its Competitive Research Programme (CRP Award No. NRF-CRP 8-2011-07). KJAO is supported by a PhD scholarship funded by the MOE Tier2 grant (2008-T2-01-033).

# References

1. R. Soref, "Mid-infrared photonics in silicon and germanium," Nat. Photonics **4**, 495-497 (2010).

2. H. Raether, Surface Plasmons on Smooth and Rough Surfaces and on Gratings (Springer, Berlin, 1988).

3. R. Standley, "Plasmonics in the mid-infrared," Nat. Photonics **6**, 409-411 (2012).

4. N. Yu and F. Capasso, "Wavefront engineering for mid-infrared and terahertz quantum cascade lasers," J. Opt. Soc. Am. B **27**, B18-B35 (2010).

5. P. Chen, Q. Gan, F. J. Bartoli, and L. Zhu, "Spoof-surface-plasmon assisted light beaming in mid-infrared," J. Opt. Soc. Am. B **27**, 685-689 (2010).

6. K. F. MacDonald and N. I. Zheludev, "Active plasmonics: current status," Laser Photon. Rev. **4,** 562–567 (2010).

7. T. Nikolajsen, K. Leosson and S. I. Bozhevolnyi, "Surface plasmon polariton based modulators and switches operating at telecom wavelengths," *Appl. Phys. Lett.* **85,** 5833 (2004).

8. M. J. Dicken, L. A. Sweatlock, D. Pacifici, H. J. Lezec, K. Bhattacharya and H. A. Atwater, "Electrooptic modulation in thin film barium titanate plasmonic interferometers," *Nano Lett.* **8,** 4048–4052 (2008).

9. A. N. Grigorenko, M. Polini and K. Novoselov, "Graphene plasmonics," *Naure Photon.* **6,** 749–758 (2012).

10. S. A. Mikhailov and K. Ziegler, "New electromagnetic mode in graphene," *Phys. Rev. Lett.* **99,** 016803 (2007).

11. M. Jablan, H. Buljan and M. Soljačić, "Plasmonics in graphene at infrared frequencies," *Phys. Rev.* B **80,** 245435 (2009)

12. A. Vakil and N. Engheta, "Transformation optics using graphene," *Science* **332,** 1291–1294 (2011).

13. C. H. Gan, H. S. Chu and E. P. Li, "Synthesis of highly confined surface plasmon modes with doped graphene sheets in the midinfrared and terahertz frequencies," *Phys. Rev.* B **85,** 125431 (2012).

14. L. Ju, B. Geng, J. Horng, C. Girit, M. Martin, Z. Hao, H. A. Bechtel, X. Liang, A. Zettl, Y. R. Shen and F. Wang, "Graphene plasmonics for tunable terahertz metamaterials," *Nature Nanotechnol.* **6,** 630–634 (2011).

15. Z. Fei, A. S. Rodin, G. O. Andreev, W. Bao, A. S. McLeod, M. Wagner, L. M. Zhang, Z. Zhao, M. Thiemens, G. Dominguez, M. M. Fogler, A. H. Castro Neto, C. N. Lau, F. Keilmann and D.

N. Basov, "Gate-tuning of graphene plasmons revealed by infrared nano-imaging," *Nature* **487,** 82–85 (2012).

16. J. Chen, M. Badioli, P. Alonso-González, S. Thongrattanasiri, F. Huth, J. Osmond, M. Spasenović, A. Centeno, A. Pesquera, P. Godignon, A. Zurutuza Elorza, N. Camara, F. Javier García de Abajo, R. Hillenbrand and F. H. L. Koppens, "Optical nano-imaging of gate-tunable graphene plasmons," *Nature* **487,** 77–81 (2012).

17. K. S. Novoselov, A. K. Geim, S. V. Morozov, D. Jiang, Y. Zhang, S. V. Dubonos, I. V. Grigorieva and A. A. Firsov, " Electric Field Effect in Atomically Thin Carbon Films," *Science* **306,** 666–669 (2004).

18. Z. Lu and W. Zhao, "Nanoscale electro-optic modulators based on graphene-slot waveguides," J. Opt. Soc. Am. B **29**, 1490-1496 (2012).

19. X. He and S. Kim, "Graphene-supported tunable waveguide structure in the terahertz regime," J. Opt. Soc. Am. B **30**, 2461-2468 (2013).

20. R. Hao, W. Du, H. Chen, X. Jin, L. Yang, and E. P. Li, "Ultra-compact optical modulator by graphene induced electro-refraction effect," Appl. Phys. Lett. **103**, 061116 (2013).

21. X. Zhu, W. Yan, N. A. Mortensen and S. Xiao, "Bends and splitters in graphene nanoribbon waveguides," *Opt. Express* **21,** 3486–3491 (2013).

22. S. Y. Zhou, D. A. Siegel, A. V. Fedorov and A. Lanzara, "Metal to insulator transition in epitaxial graphene induced by molecular doping," *Phys. Rev. Lett.* **101,** 086402 (2008).

23. Z. Q. Li, E. A. Henriksen, Z. Jiang, Z. Hao, M. C. Martin, P. Kim, H. L. Stormer and D. N. Basov, "Dirac charge dynamics in graphene by infrared spectroscopy," *Nature Phys.* **4,** 532–535 (2008).

24. H. Liu, Y. Liu and D. Zhu, "Chemical doping of graphene," *J. Mater. Chem.* **21,** 3335–3345 (2011).

25. P. Berini, "Plasmon-polariton waves guided by thin lossy metal films of finite width: Bound modes of symmetric structures," *Phys. Rev.* B **61,** 10484–10503 (2000).

26. A. Yu. Nikitin, F. Guinea, F. J. Garcia-Vidal, and L. Martin-Moreno, "Edge and waveguide terahertz surface plasmon modes in graphene microribbons," *Phys. Rev.* B **84**, 161407(R) (2011).

27. J. M. Foley, A. M. Itsuno, T. Das, S. Velicu and J. D. Phillips, "Broadband long-wavelength infrared Si/SiO$_2$ subwavelength grating reflector," *Opt. Lett.* **37,** 1523–1525 (2012).

28. D. R. Andersen, "Graphene-based long-wave infrared TM surface plasmon modulator," *J. Opt. Soc. Am. B* **27,** 818–823 (2010).

29. E. Feigenbaum and M. Orenstein, "Perfect 4-way splitting in nano plasmonic X-junctions," *Opt. Express* **15,** 17948–17953 (2007)

30. P. Bai, M. X. Gu, X. C. Wei and E. P. Li, "Electrical detection of plasmonic waves using an ultra-compact structure via a nanocavity," *Opt. Express* **17,** 24349–24357 (2009).

31. A. Pannipitiya, I. D. Rukhlenko, M. Premaratne, H. T. Hattori and G. P. Agrawal, "Improved transmission model for metal-dielectric-metal plasmonic waveguides with stub structure," *Opt. Express* **18,** 6191–6204 (2010).

32. E. Feigenbaum and H. A. Atwater, "Resonant guided wave networks," *Phys. Rev. Lett.* **104,** 147402 (2010).

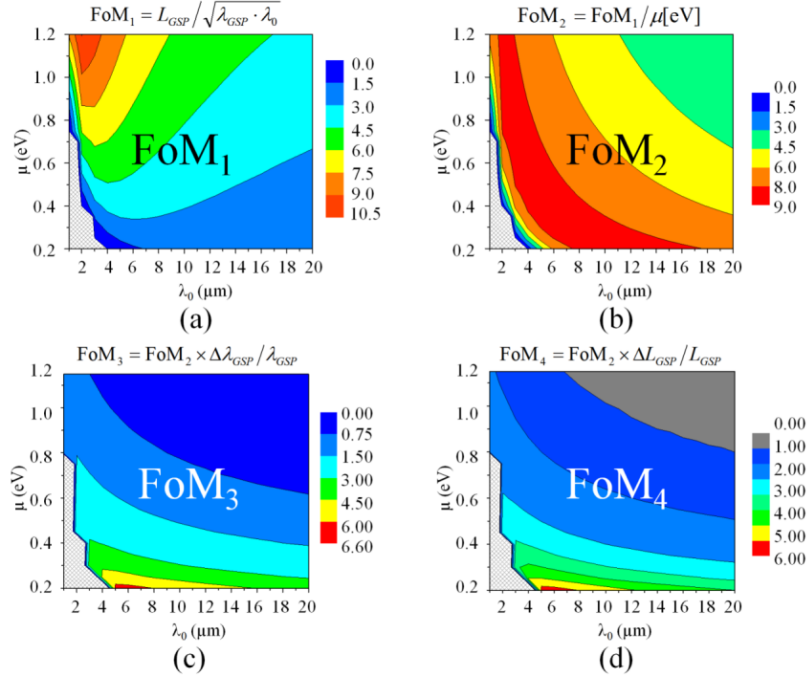

Figure 1: Figure-of-Merit (FoM) contour plots for passive and active waveguiding.

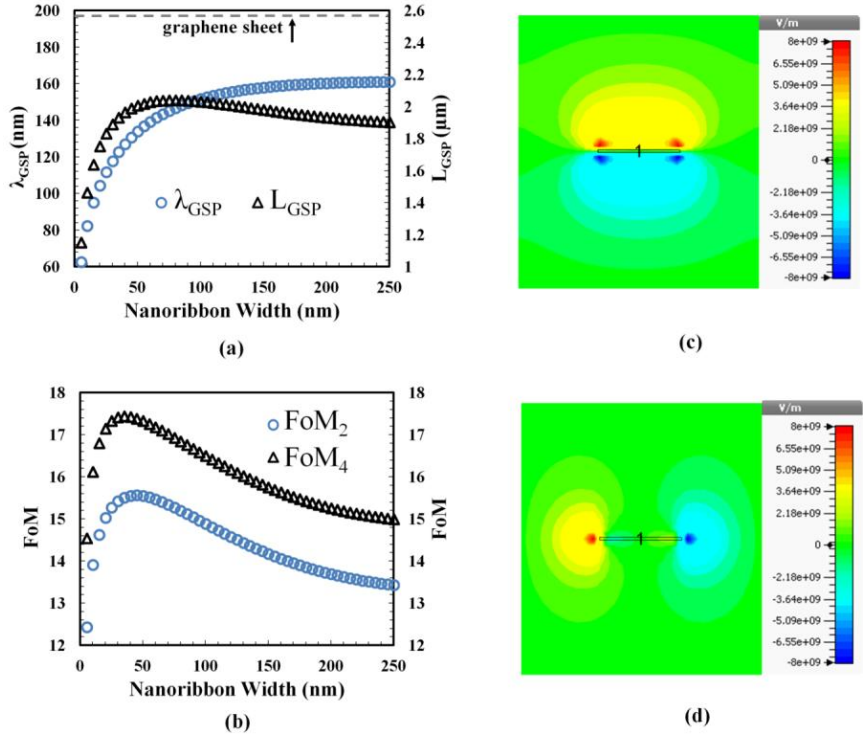

Figure 2: (a) Propagation characteristics by means of the effective wavelength $\lambda_{GSP}$ and propagation loss $L_{GSP}$ as a function of nanoribbon widths. (b) Figure-of-Merits $FoM_2$ and $FoM_4$ as a function of nanoribbon widths. (c) Z-direction and (d) y-direction electric-field component of the fundamental propagation mode of a 30nm-wide GNR.

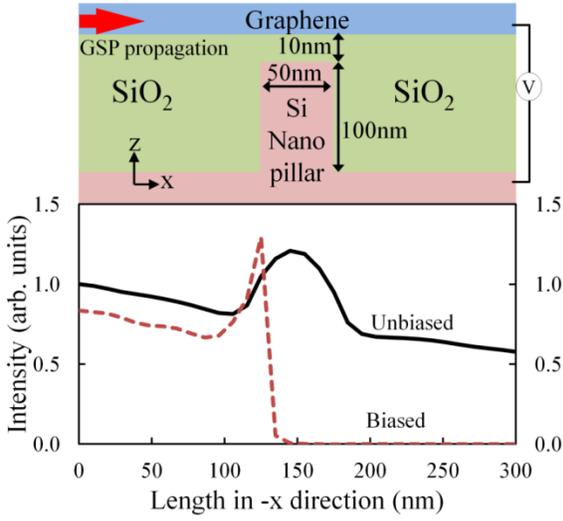

Figure 3: Graphene plasmonic shutter employing a GNR on an inhomogeneous dielectric layer. Applying a gate-voltage can modulate the GSP transmission by >30dB, as shown in the intensity plot.

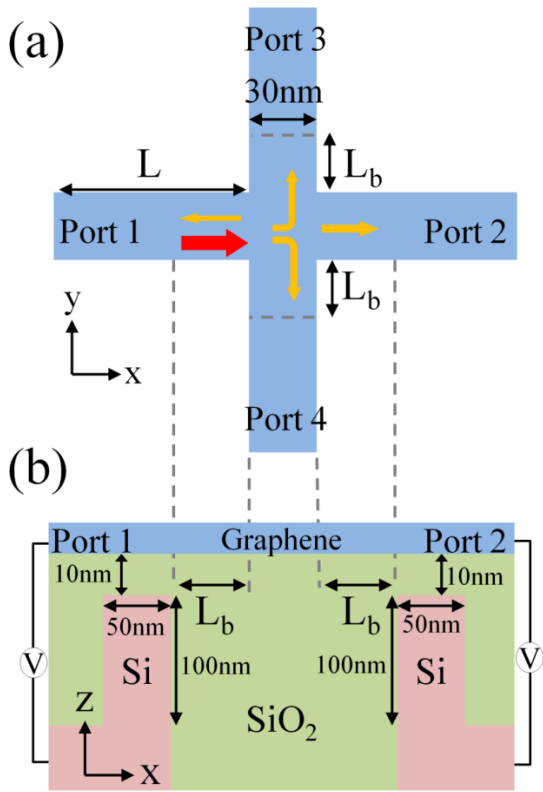

Figure 4: (a) Top view of a 4-port GNR cross-junction waveguide. (b) Horizontal cross-section from Port 1 to Port 2 of the GNR cross-junction waveguide. The Si nanopillars are electrically isolated, separated by a total buffer-zone length and junction width of $2\times L_b+30$nm.

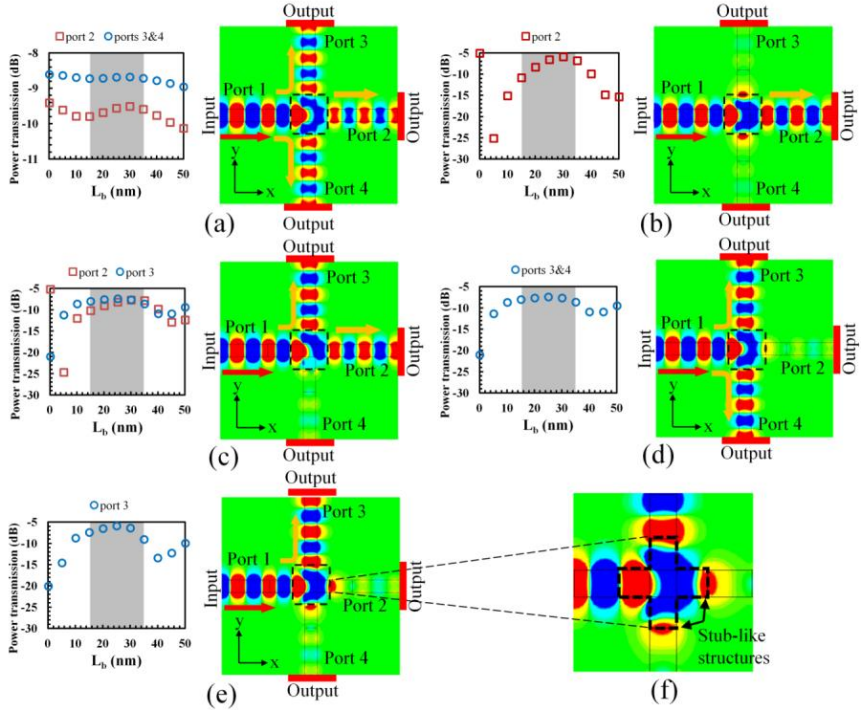

Figure 5: Power transmission at the output ports as a function of buffer length $L_b$ and their corresponding z-direction electric-field maps of GSP propagation on the cross-junction waveguide, with transmission switched on for (a) all output ports (unbiased-state), (b) Port 2 only, (c) Ports 2 and 3, (d) Ports 3 and 4, and (e) Port 3 only. The buffer zones are indicated in the electric-field maps in dotted boxes. Electric-field map of buffer zone for case (e) is enlarged in (f) to clearly illustrate the formation of the stub-like structures when two waveguide arms are switched off.

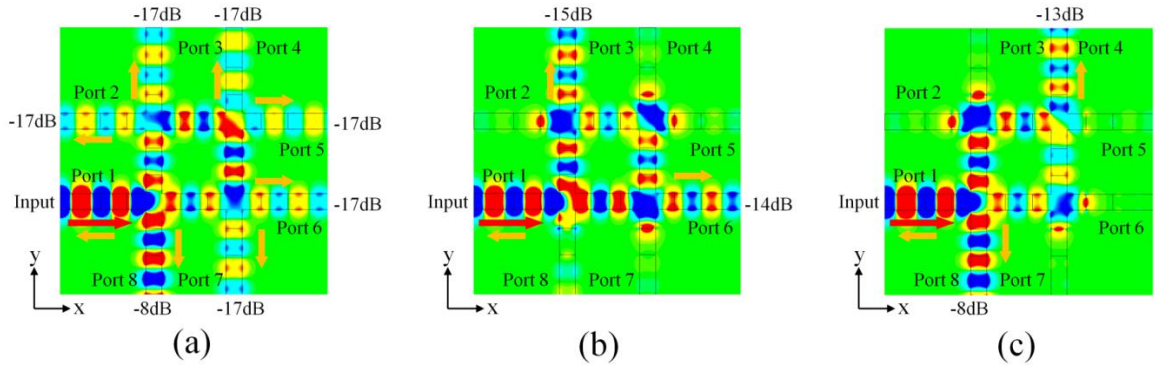

Figure 6: Z-direction electric-field maps of GSP propagation on an 8-port GNR waveguide network, with transmission switched on for (a) all output ports, (b) Ports 3 and 6, and (c) Ports 4 and 8. Respective port transmission intensities are indicated.